\documentclass[preprint2]{aastex631} 

\shorttitle{Particle Acceleration in Reconnection with Scattering}
\shortauthors{JOHNSON ET AL.}
\usepackage{amsmath}
\DeclareMathOperator{\sech}{sech}
\usepackage{hyperref}

\begin{document}

\title{Particle Acceleration in Magnetic Reconnection with Ad hoc Pitch-angle Scattering}

\correspondingauthor{Grant Johnson}
\email{grj@princeton.edu}

\author[0000-0002-5981-4688]{Grant Johnson}
\affiliation{Princeton Plasma Physics Laboratory \\
100 Stellarator Rd, Princeton, NJ 08540}

\author[0000-0002-8906-7783]{Patrick Kilian}
\affiliation{Space Science Institute \\
 4765 Walnut St, Suite B, Boulder, CO 80301}

\author[0000-0003-4315-3755]{Fan Guo}
\affiliation{Los Alamos National Laboratory\\
Los Alamos, NM 87545, USA}
\affiliation{New Mexico Consortium\\
4200 West Jemez Rd, Los Alamos, NM 87544, USA}

\author[0000-0001-5278-8029]{Xiaocan Li}
\affiliation{Dartmouth College\\
Hanover, NH 03755}

\begin{abstract}

Particle acceleration during magnetic reconnection is a long-standing topic in space, solar and astrophysical plasmas. Recent 3D particle-in-cell simulations of magnetic reconnection show that particles can leave flux ropes due to 3D field-line chaos, allowing particles to access additional acceleration sites, gain more energy through Fermi acceleration, and develop a power-law energy distribution. This 3D effect does not exist in traditional 2D simulations, where particles are artificially confined to magnetic islands due to their restricted motions across field lines. Full 3D simulations, however, are prohibitively expensive for most studies. Here, we attempt to reproduce 3D results in 2D simulations by introducing ad hoc pitch-angle scattering to a small fraction of the particles. We show that scattered particles are able to transport out of 2D islands and achieve more efficient Fermi acceleration, leading to a significant increase of energetic particle flux. We also study how the scattering frequency influences the nonthermal particle spectra. This study helps achieve a complete picture of particle acceleration in magnetic reconnection.

\end{abstract}

\section{Introduction} \label{sec:Introduction}
Magnetic reconnection converts magnetic energy into plasma energies by releasing free energy stored in the configuration of field lines of force \citep{Yamada2010,Ji2022}. It has been proposed as a mechanism for producing high energy charged particles in space, solar, and astrophysical processes. For example, evidence of energetic particle production has been found in Earth's magnetotail \citep{Oieroset2002,Fu2011,Fu2013,Ergun2018} and solar flares \citep{Su2013,Oka2015,Chen2018,Chen2020}. The nonthermal acceleration has also been invoked to explain high energy particle acceleration and radiation in astrophysical systems such as pulsar winds and relativistic jets from black holes and neutron stars \citep[See][and references therein]{Guo2020}. 

Although particle acceleration during magnetic reconnection has been studied for several decades, many aspects are still unclear. Recent studies suggest that Fermi acceleration by contracting or merging magnetic islands is the main acceleration mechanism during magnetic reconnection \citep{Drake2006,Dahlin2014,Guo2014,Guo2015,Guo2019,Li2017,Li2018,Li2019b,Zank2014,leRoux2015,Wang2016,Du2018,Kilian2020}. This mechanism can produce a large number of energetic particles in the reconnection layer filled with magnetic islands and is promising for explaining efficient particle acceleration in solar flares~\citep{Li2018Large,Arnold2021PRL,Zhang2021}. One particular concern was that early kinetic simulations had difficulties producing power-law energy distributions \citep[e.g.,][]{Drake2010,Dahlin2014,Li2017,Che2020}, a common feature of energetic particle measurements, despite various attempts \citep[see summaries by][]{Guo2020,Li2021}. Several recent studies have been devoted to the formation of power-law energy spectrum in nonrelativistic reconnection conditions. First, the development of nonthermal processes happens preferentially at the low plasma $\beta$, low guide field condition, where particles are accelerated quickly so that the accelerated nonthermal distribution may outgrow the background thermal distribution \citep{Li2015,Li2017,Li2019}. Second, in 2D particle-in-cell (PIC) simulations, particles are artificially confined inside magnetic islands due to restricted motions across field lines within $\sim$ one gyroradius\footnote{As noted by \citet{Jokipii1993} and \citet{Jones1998}, the restricted cross-field motion is due to the conservation of canonical momentum when at least one ignorable coordinate exists.}. High-energy particles cannot be further accelerated and therefore the nonthermal distributions in 2D simulations are transient \citep{Li2017,Li2019}. Recent work in the nonrelativistic regime suggests that 3D effects can facilitate the acceleration of particles \citep{Dahlin2015,Dahlin2017,Li2019,Zhang2021}. In this scenario, particles can leave their original flux rope due to field-line chaos, which leads to more efficient particle transport in the reconnection layer. Consequently, particles can access multiple acceleration regions and gain more energy through extra Fermi bounces. This leads to a nearly constant acceleration rate as a function of energy and facilitates the development of nonthermal distributions \citep{Li2019,Zhang2021}. The resulting 3D energy spectra are harder, and the high-energy cutoff is greater than those in 2D simulations. Although 3D kinetic simulations can capture the 3D effects self-consistently, they demand a tremendous amount of computational resources, prohibiting us from conducting a systematic study.

Since the restricted particle motion across field lines in 2D systems is intrinsically due to the charged particle Hamiltonian in an electromagnetic field, it is desirable to develop techniques to remove this restriction, so that 2D simulations reproduce 3D results. Here in 2D PIC simulations we introduce ad hoc pitch-angle scattering which randomizes the momentum directions of the particles, thus breaking the invariant. 
We explore if power-law spectra similar to those in the 3D simulations can be reproduced in 2D simulations with ad hoc pitch-angle scattering. This is meant to be analogous to the cross-field transport seen in 3D reconnection simulations~\citep{Dahlin2015,Li2019,Zhang2021,Guo20213D}. We will show that by allowing the particles to move across field lines, beyond that of the gyromotions, they are able to access more acceleration sites and statistically gain more energy. This affects the energy spectrum, hardening the 2D spectrum and increasing the number of particles in the high-energy tail. While our focus will be on electron spectra, we will also present the results of scattered ions.

The paper is laid out as follows. Section \ref{sec:VPIC simulations} will elaborate on the parameters used for our PIC simulations and describe the scattering model. Then, the results of the simulations, and a comparison between the 2D and 3D results, are given in Section \ref{sec:Discussion}. Concluding remarks and future directions for research are then presented in Section \ref{sec:Conclusion}.

\section{Numerical simulations} \label{sec:VPIC simulations}

The numerical simulations are carried out using the VPIC code that solves the Maxwell-Vlasov system of equations using the PIC method \citep{Bowers2008}.
We initially setup a force-free current sheet with magnetic field $\mathbf{B} = B_0 \tanh{(z/L)}\hat{x} + B_0\sqrt{\sech{(z/L)}^2 + b_g^2}\hat{y}$ and a small perturbation to the field in the $x-z$ plane~\citep{Birn2001Geospace}. The main physical parameters are ion and electron temperatures $T_i = T_e$ = 0.01 $m_{e}c^2 / k_B$, ratio of ion to electron mass $m_i/m_e$ = 25, reconnecting layer characteristic-width $L$ = 1$d_i$ (ion inertial length), a guide field ratio $b_g = B_g/B_0$ = 0.2, and ratio of the electron plasma frequency to the electron cyclotron frequency $\omega_{pe}/\Omega_{ce}$ = 1.0. The initial current density, given by Ampere's law, is carried by the electron population. We used the following quantities to normalize the simulation results: the speed of light $c$, reference magnetic field strength $B_0$, electron mass $m_e$, and the ion inertial length $d_i$. A typical simulation proceeds until  $t\Omega_{ci} = 400$ ($\Omega_{ci}$ is the ion-cyclotron frequency). The box-size for most simulations is $L_x = 150d_i$ and $L_z = 62.5d_i$ on a grid of $n_x \times n_z = 3072 \times 1280$ cells. All simulations were run with a timestep of 0.8 times the CFL limit ($\Delta{t} = 0.8\Delta{x}/\sqrt{2}c$) on a uniformly spaced 2D mesh with $\Delta{x} = L_x/n_x = L_z/n_z$. These simulations parameters where consciously chosen to match the parameters in recent 3D simulations \citep{Li2019,Zhang2021} for comparison between the 2D and 3D simulations. To match these 3D simulations we used 150 particles per cell per species. Our boundaries are periodic in $x$ for both fields and particles and reflective to particle in $z$ with perfectly conducting boundaries for fields. The 3D simulations from \citet{Li2019} have additionally $n_y$ = 1536 ($L_y = 75d_i$). Lastly, for comparing the ion spectra with a recent study by \citet{Zhang2021}, we double the domain size and number of cells ($L_x = 300d_i$ and $L_z = 125d_i$ and $n_x \times n_z = 6144 \times 2560$) while keeping all other parameters the same.

In addition to the bulk ion and electron species, we created two additional electron species: one undergoing scattering and the other without scattering. These species were allowed to feedback on the simulation (contributing to the current density and charge density) and consisted of 0.5\% of the total electron population for each of the two additional species. Henceforth these two additional electron groups are noted as ``scattered" and ``unscattered", respectively. The remaining electron species will be referred to as the bulk species.  By partitioning these additional species from the total species and keeping the same weight of the particles, we preserve global charge neutrality. We examine the effects of scattering primarily through studying the particle energy spectra and distributions in space, and comparison with the bulk particle spectra in 2D and 3D. As a final important technical note, we find that particle feedback effect is important in these low-$\beta$ simulations so that particles have negligible numerical heating.

\subsection{Scattering Model}
As the first study on the role of ad hoc pitch-angle scattering in magnetic reconnection, we implement a simple model of scattering. We randomize the pitch angle in the local plasma flow frame ($\bold{v}_{f} = \sum_{s} \rho_{s}\langle \bold{v}_{s} \rangle/\sum_{s} \rho_{s}$) where $\rho_{s}$ and $\langle \bold{v}_{s} \rangle$ are the local mass density and bulk velocity vector, respectively, of the species $s$. The particle energy in the flow frame is conserved. However, momentum and energy are not conserved in the simulation rest frame. Similar models have been used to study particle acceleration at collisionless shocks \citep[e.g.,][]{Giacalone1994}. We confirmed that scattering does not significantly contribute to the total energy of the simulation, and particle energy gain during scattering is negligible. Instead, the change of energy is caused primarily by scattered particles accessing more acceleration sites.

Individual particles are scattered with probability $\chi = \nu \Delta{t}$ that is determined by the scattering frequency $\nu$ during a single time step $\Delta{t}$.
For the following pitch-angle scattering algorithm, $\mathbf{v}$ is the particle velocity and $\mathbf{u} = \gamma \mathbf{v}$, where $\gamma$ is the Lorentz factor and $\mathbf{u}$ the spatial components of the four-velocity; $\mathbf{u}_f$ and $\gamma_f$ are the spatial components of the four-velocity of the plasma flow velocity and its Lorentz factor, respectively. 
The first step in scattering the particle is to transform $\mathbf{u}$ to the local plasma flow frame, which are denoted as the primed coordinates. Starting with the four-velocity in the simulation frame (unprimed coordinates), we can calculate the particle four-velocity in the flow frame, from Equation 6b in \citet{strel1992addition},

\begin{equation} \label{eq:velocity_addition}
    \mathbf{u}' = \mathbf{u} - \mathbf{u}_f\frac{\gamma + \gamma\gamma_f - \mathbf{u} \cdot \mathbf{u}_f/c^2}{1+ \gamma_f}
\end{equation}
For the next step, the particle is scattered by randomly reorienting the direction of $\mathbf{u'}$ to a new direction $\mathbf{u}''$ without changing its magnitude $u' = |\mathbf{u}'|$. This is accomplished by choosing two uniform random numbers $R_1 \in (0,1)$ and  $R_2 \in (0,2\pi)$ together with
\begin{equation}
\begin{split}
    &\mathbf{u}_x'' =  u'\sqrt{1-R_1^2}\cos(R_2)\\
    &\mathbf{u}_y'' =  u'\sqrt{1-R_1^2}\sin(R_2)\\
    &\mathbf{u}_z'' =  u'R_1.
\end{split}
\end{equation}

Finally, by reversing the sign of $\mathbf{u}_f$ in Equation~\ref{eq:velocity_addition}, we can transform the velocity back to the simulation rest frame. Once the new velocity is calculated, the scattering is complete.

The ad hoc scattering breaks the invariant $P_y = p_y + qA_y/c$, by randomizing the magnitude of $p_y$ while keeping the vector potential at the particle location $A_y$ constant, and thus remove the particles' tie to the field line similar to the 3D reconnection processes \citep{Dahlin2017,Li2019,Zhang2021}.

\section{Simulation Results} \label{sec:Discussion}

\begin{figure*}[ht!]
\includegraphics[width=\linewidth]{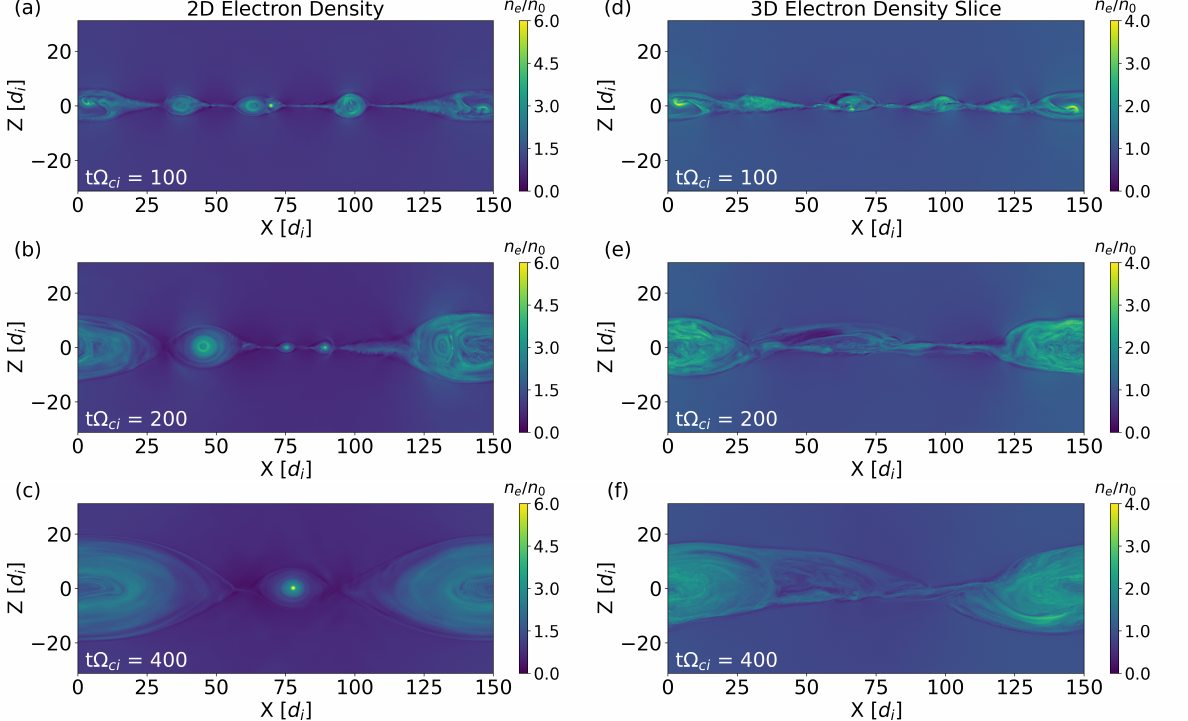}
\caption{Comparison of the normalized 2D to 3D electron densities $n_\mathrm{e} / n_\mathrm{0}$ at three time steps. The 3D data is taken over a single slice at $y = 0$.}
\label{fig:density_profile}
\end{figure*}

To highlight the differences between 2D and 3D reconnection, Figure~\ref{fig:density_profile} shows the electron density distributions in 2D and 3D simulations. In both simulations, the current layer breaks into a series of magnetic islands (or flux ropes) as the reconnection begins. In the 2D simulation (left panels), the islands merge to form larger islands, and secondary islands are continuously generated in the thin reconnection layer. Electrons tend to concentrate in the center of the islands during these processes, and the resulting electron density is much higher therein. In contrast, electrons in 3D reconnection are less concentrated in flux ropes due to 3D dynamics, and the electron distributions are more uniform (right panels) than those in 2D.

\begin{figure*}[ht!]
\centering
\includegraphics[width=\linewidth]{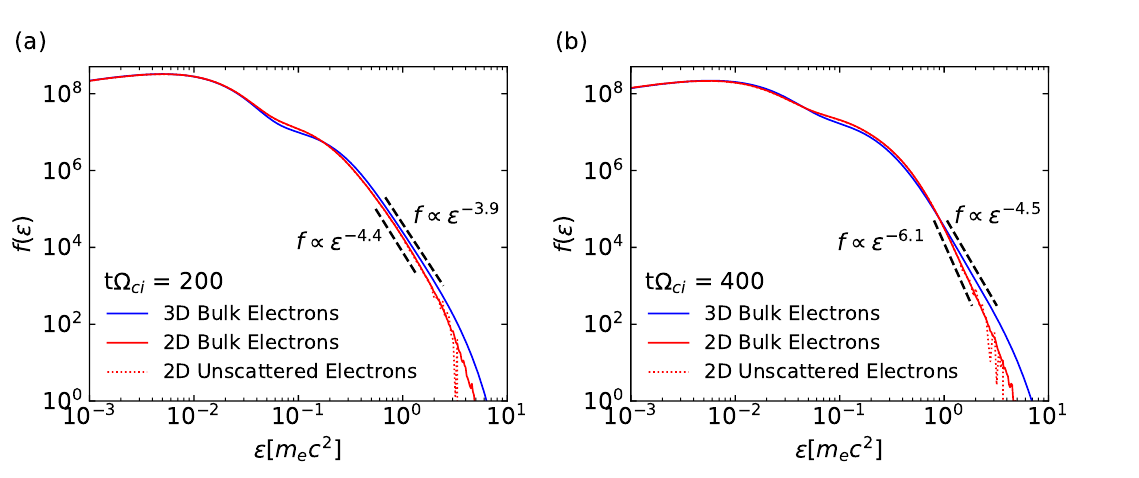}
\caption{Comparison of the 3D electron spectra~\citep[adapted from][]{Li2019} to the 2D bulk electron spectrum and the unscattered electron species at the same parameters. The panels are taken at two time slices of $t\Omega_{ci} = 200$ (a) and $t\Omega_{ci} = 400$ (b). The dashed black lines indicate the power-law spectral slopes. The spectra have a bulk of the distribution located at $\varepsilon \sim k_BT_e/m_ec^2$ since the spectra are integrated over the entire simulation domain, where the majority of the particles remain upstream in the thermal population.
\label{fig:2D_v_3D_spectrum}}
\end{figure*}

\begin{figure*}[ht!]
\centering
\includegraphics[width=\linewidth]{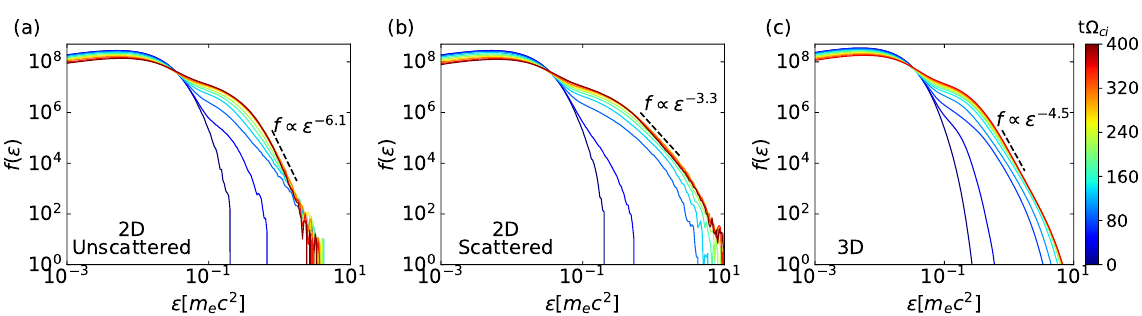}
\caption{Time evolution of the electron energy spectra from time 0 (dark blue) to $t\Omega_{ci} = 400$ (dark red). (a) The spectrum of the unscattered electrons in the 2D simulation. (b) The spectrum of the scattered electrons with a scattering frequency of $\nu = 1.0\Omega_{ci}$ in the 2D simulation. (c) The spectrum in the corresponding 3D simulation (adapted from~\citet{Li2019}). Note that the spectra have been normalized so that the total electron densities are the same.
\label{fig:electron_spec_compare}}
\end{figure*}

In 3D, fast transport of energetic electrons enables them to access more acceleration regions and gain higher energies through the Fermi mechanism~\citep{Dahlin2015,Li2019,Zhang2021}. Consequently, the electron energy distributions evolve differently in 2D and 3D simulations. Figure~\ref{fig:2D_v_3D_spectrum} compares the electron energy distributions in the two simulations shown in Figure~\ref{fig:density_profile}. The spectra in the 2D simulation are for bulk electrons and unscattered electron species. We reassure the similarity of the unscattered species with the bulk species since we will use data from the unscattered species to contrast these with the scattered species in subsequent figures. To reduce the noise, we average the scattered and unscattered electron spectra over $ \sim 0.5\Omega_{ci}^{-1}$. Overall, Figure \ref{fig:2D_v_3D_spectrum} shows that electrons are accelerated to higher energy, and the spectra are harder in the 3D simulation than in 2D. In both simulations, the high-energy part of the distributions resembles a power law $f(\varepsilon)\sim\varepsilon^{-p}$. At $t\Omega_{ci} = 200$, the power-law index $p\approx3.9$ in the 3D simulation and about 4.4 in the 2D simulation. As the simulation proceeds, the spectra in both simulations become softer due to the periodic boundary conditions, which slow down the reconnection outflows and thus the Fermi acceleration due to the motional electric field. At $t\Omega_{ci} = 400$, $p\approx4.5$ in the 3D simulation and about 6.1 in the 2D simulation. The discrepancy in the spectral index persists after about $t\Omega_{ci} = 200$ and becomes broader due to the artificial trapping by magnetic islands in 2D, as we will further show in Figure \ref{fig:electron_spec_compare}(a). This is caused by suppression of the high-energy particle acceleration. We expect particle scattering in 2D will prevent electrons from being trapped in the islands, enhance electron acceleration, and minimize the discrepancy between 2D and 3D simulations.

Figure \ref{fig:electron_spec_compare} compares the spectra of three different electron populations including the species that experiences no scattering (``unscattered'') in 2D, and with scattering (``scattered'', $\nu = 1.0\Omega_{ci}$) in 2D, and the bulk electron population in the 3D simulation \citep{Li2019}. All species start with the same initial distribution, and each forms a power-law spectrum. Besides the different spectral slopes highlighted in  Figure~\ref{fig:2D_v_3D_spectrum}, the time evolution reveals more differences between 2D and 3D simulations. Figure~\ref{fig:electron_spec_compare}(a) shows that high-energy particle acceleration ($\varepsilon \gtrsim m_ec^2$) effectively ceased around $t\Omega_{ci} = 200$ in 2D, resulting a spectrum softer and softer over time. In contrast, Figure~\ref{fig:electron_spec_compare}(c) shows that the acceleration is sustainable in 3D with a continuous increase in the maximum energy.
Towards the end of the simulations, the scattered electrons (Figure~\ref{fig:electron_spec_compare}(b)) are accelerated to higher energies than the other two species and develop the hardest power-law energy spectrum among the three species ($p=3.3$). This result demonstrates that the scattering can enhance electron acceleration and increase high-energy electron fluxes in 2D simulations. However, the acceleration of the scattered species is even stronger than that in the 3D simulation, suggesting that the scattering might be too frequent. 

\begin{figure*}[ht!]
\centering
\includegraphics[width=\linewidth]{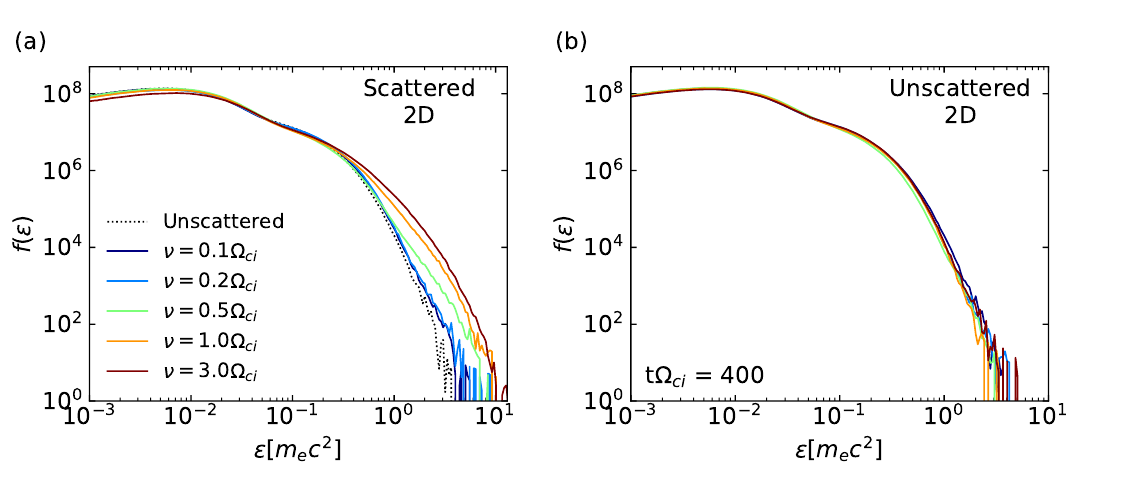}
\caption{A scan over different scattering frequencies ($\nu$) in 2D. (a) The spectra of the scattered electrons at $t\Omega_{ci} = 400$. The dashed line shows the spectrum of the unscattered electrons in the $\nu = 0.1\Omega_{ci}$  simulation for reference. (b) The spectra of the corresponding unscattered electrons for the same simulations.
\label{fig:scan_spectrum_compare}}
\end{figure*}

To better understand how the scattering frequency influences the particle spectra, we performed a scan of the scattering frequencies. Figure \ref{fig:scan_spectrum_compare} shows the results of this scan over a range of relevant frequencies $\nu =  0.1 - 3.0 \Omega_{ci}$. Panel a) shows the spectra of the scattered species. When $\nu$ is small, the spectrum is only slightly harder, and the maximum energy only marginally increases from the unscattered species spectra. However, as the scattering frequency increases, the spectral index continues to decrease. Interestingly, the spectra become curved at around a scattering frequency of 3.0 $\Omega_{ci}$, indicating too frequent scattering may hinder the acceleration. In contrast, panel (b) shows only slight variability in the unscattered electron spectra in corresponding runs, likely due to slightly different initial conditions (statistical noise), which could change the detailed reconnection dynamics and total energy converted from fields to the particles. 

Since past simulations have had difficulty producing power-law spectra, it is worth noting here that our simulations meet the conditions laid out by \cite{Li2015} to form a power-law. First, there needs to be sufficient spatial extent of the simulation to facilitate the magnetic reconnection for a long time such the particle injection time is large. The second requirement is to have a low-$\beta$ (or high Alfv\'en speed $V_A$) such that the acceleration is sufficiently fast. In other words the product of acceleration rate and injection time needs to be fairly large $\alpha \tau_{inj} > 1$ \citep[see also][]{Guo2014}. In addition, the inclusion of ad hoc pitch angle scattering with a scattering frequency $\nu < \Omega_{ci}$ allows the power-law to persist similar to earlier 3D simulations \citep{Li2019,Zhang2021}, and even make it hardened as the scattering frequency increases. 

\begin{figure*}[ht!]
\centering
\includegraphics[width=\linewidth]{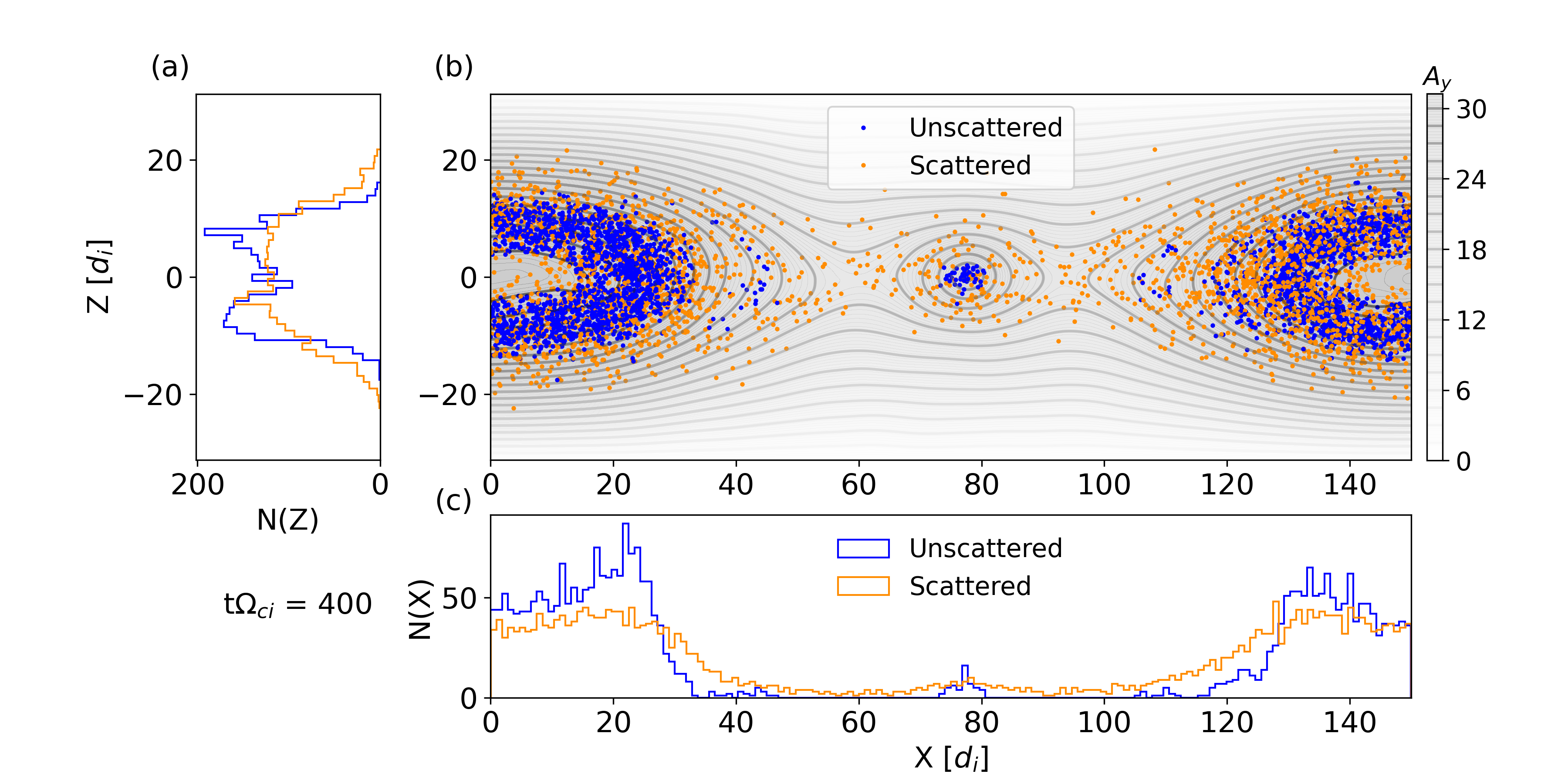}
\caption{Locations of unscattered and scattered particles with energies between 0.7 and 0.8$m_ec^2$ in the 2D simulation. The grey lines are the contours of the $y-$component of the magnetic vector potential $A_y$. The left panel (a) shows the distributions of the particles along $z$. The bottom panel (c) show the distributions of the particles along $x$.
\label{fig:tracer_versus_scattered}}
\end{figure*}

\begin{figure}[ht!]
\centering
\includegraphics[width=\linewidth]{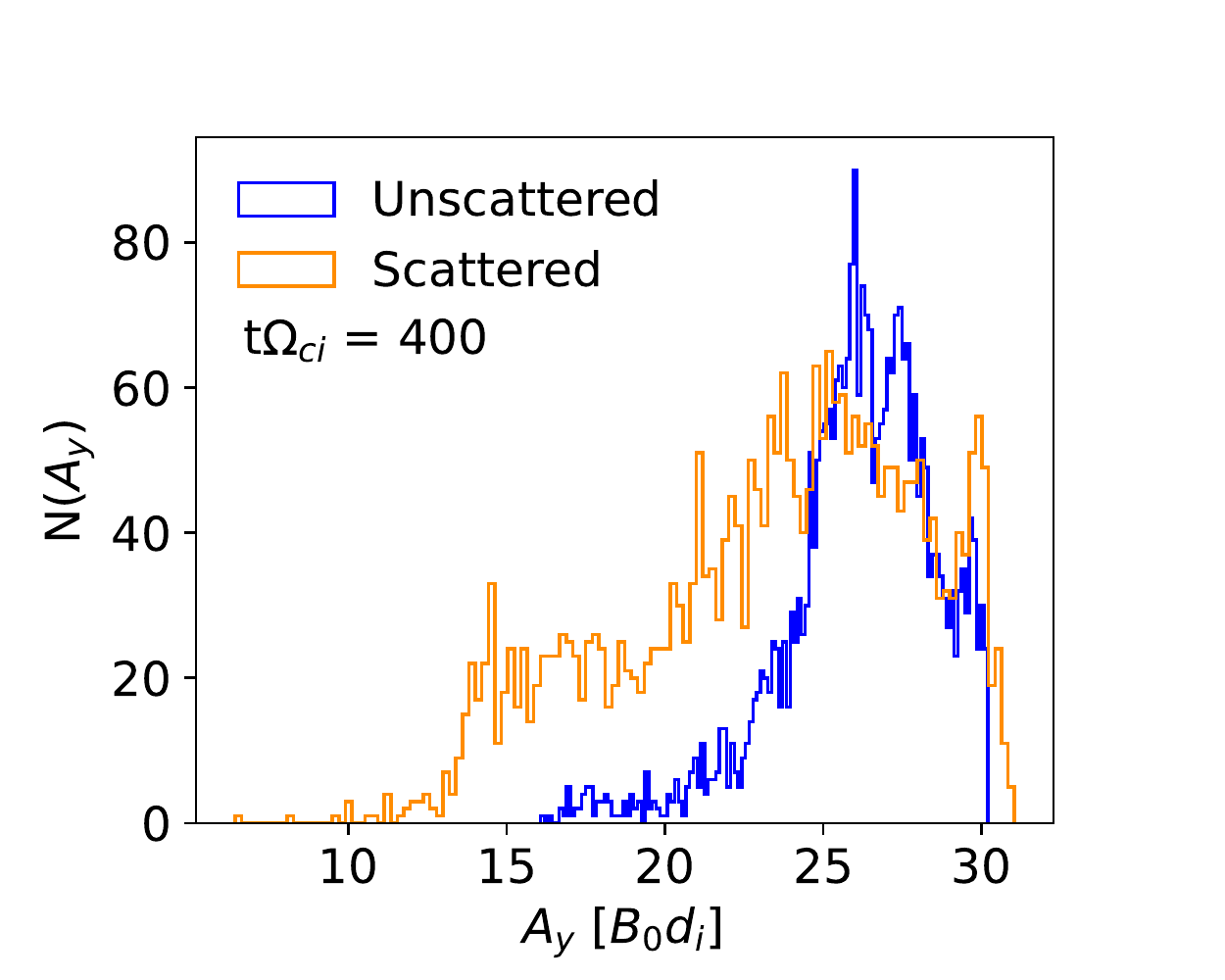}
\caption{Histograms of the scattered and unscattered electrons within the energy range between 0.7 and 0.8$m_ec^2$ corresponding to different values of $A_y$, using the same population as those in Figure \ref{fig:tracer_versus_scattered}.
\label{fig:Ay_hist}}
\end{figure}

The ability for scattered particles to leave the islands is examined next in Figure \ref{fig:tracer_versus_scattered}. The large panel (b) shows the magnetic vector potential overlaid with 3000 unscattered and scattered particles in the energy range $0.7-0.8m_ec^{2}$ (within the power-law ranges of both scattered and unscattered electrons) for the case of $\nu = 1.0 \Omega_{ci}$. The blue unscattered electrons are well confined within magnetic islands as they are tied to field lines with their original $A_y$. The scattered electrons, however, are able to leave the islands and fill the exhaust regions, where the Fermi acceleration is most active,  as we have seen in 3D simulations \citep[e.g.,][]{Li2019}. Some of the scattered electrons can even reach the reconnection inflow region, similar to the observed particle diffusion across the exhaust boundaries in 3D simulations~\citep{Le2018Drift}.
We emphasise the differences in panels (a) and (c) which show the distribution of particles in $z$ and $x$, respectively. The spread of particles in space is apparent around the edges of the islands, where spatial diffusion due to scattering has flattened the peaks of the scattered electrons relative to those of the unscattered electrons. Panels (a) and (b) of Figure \ref{fig:tracer_versus_scattered} demonstrates that energetic, scattered particles are able to leave the islands while remain primarily within the current layer. Another way to look at this diffusive process is through Figure \ref{fig:Ay_hist}, which shows that the scattered particles move in the magnetic vector potential space away from the islands compared to the unscattered particles, even though the particles start with similar initial distributions. 

\begin{figure*}[ht!]
\centering
\includegraphics[width=\linewidth]{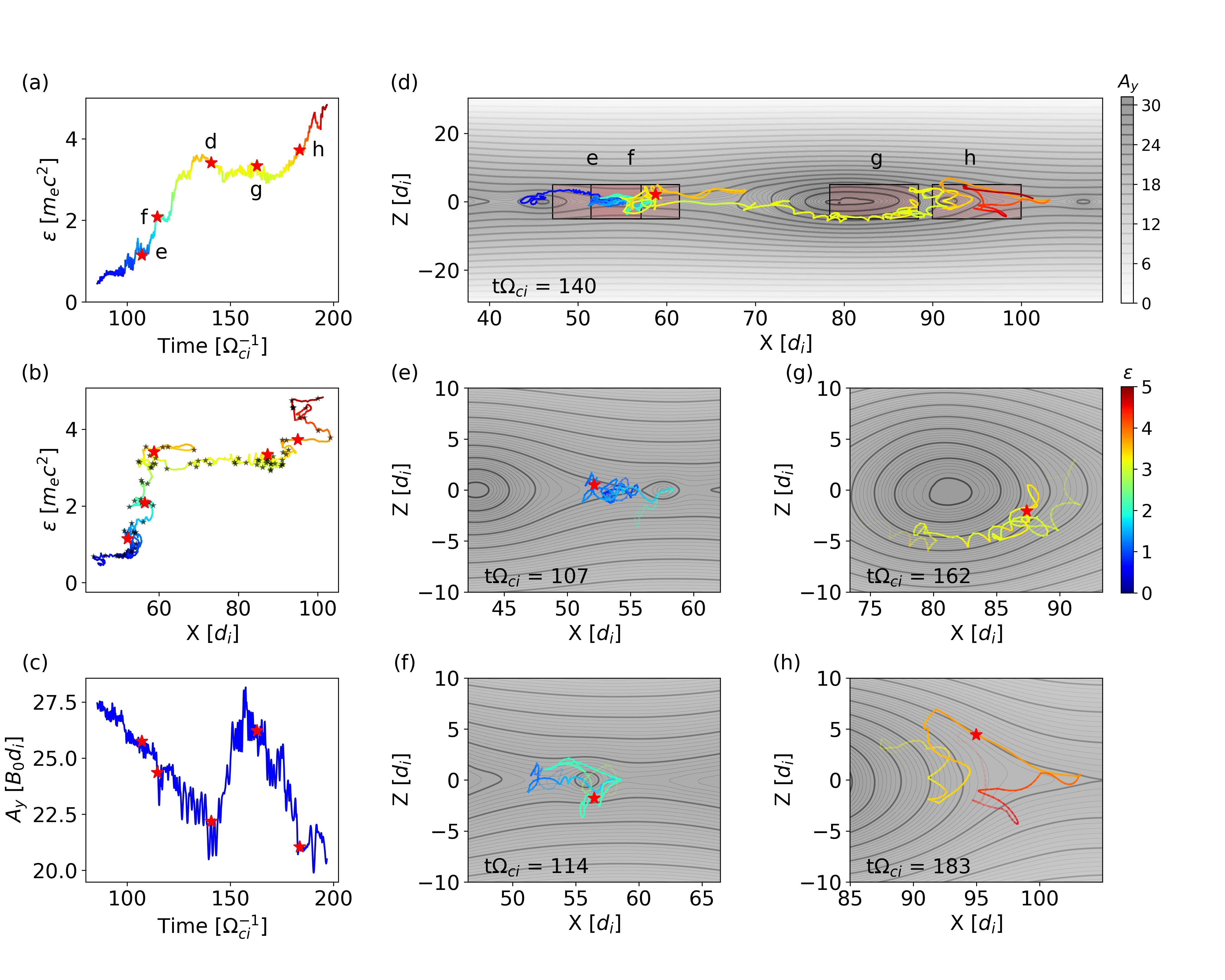}
\caption{The trajectory of one scattered particle in the 2D simulation. (a) Time evolution of the particle energy. The red stars indicate the particle location when plotted in panels (d-h) and the color indicates the kinetic energy, $\varepsilon$. (b) Particle energy, $\varepsilon$ versus its $x-$position with black stars indicating the locations where the particle undergoes scattering. (c) Time evolution of $A_y$ at the particle location. (d) Particle trajectory in the $x-z$ plane. The shaded boxes indicate the zoomed-in segments shown in panels e-h. Overplotted are the contours of $A_y$ at $t\Omega_{ci}=140$ when the electron is at the red star. (e)-(h) The segments of the particle trajectory in the shaded boxes in panel (d). The red stars indicate the particle location at the corresponding timestamps in each panel.
\label{fig:particle_trajectory}}
\end{figure*}

Examining how particles gain more energy when they undergo scattering, we find two scenarios that make particle acceleration more efficient. The first one is that some particles are transported to other islands. The other scenario is that some particles can be scattered back to the outskirts of islands (the exhaust region) and get more acceleration. In Figure \ref{fig:particle_trajectory} we show both via the trajectory of a representative high energy particle. Panel (a) shows that the particle is accelerated to approximately 5.0 times its rest energy by $t\Omega_{ci}=200$. We concentrate on the interval between energies of 0.5 and 5.0 $m_ec^2$. Crucial acceleration points are labeled with red stars and correspond to panels (d-h) that show particle trajectories in the reconnection layer. Panel (b) shows particle energy as a function of the $x$-position, where the black stars denote the time steps where the particle underwent scattering. In panel (c) we show the evolution of $A_y$ at the particle location. Due to scattering, the particle is able to go across field lines, represented in the large range of changes in $A_y$ that the particle sees. Noticeably, the particle can move toward smaller $A_y$ (outer layer of the reconnection region) and larger $A_y$ (reentering islands). 
In panel (d-h), we overlay particle trajectory on $A_y$ to show the motion of the particle in the reconnection layer. An associated animation \footnote{\url{https://youtu.be/rwqIQ5mczTY}} is available for more details. Note that in each panel, $A_y$ and its contours are plotted for the same time when the particle location is at the red star. From the overview plot panel (d), we see that this particle starts at low energy around $x \sim 45d_i$ and gains energy as it interacts with three distinct islands at $x \sim 45 d_i$, $55 d_i$  and $80d_i$, respectively. For the period in panel (e), the particle bounces a few times in the exhaust region of the first island, and move to the second small island at $x \sim 57 d_i$ at this time. During the phase around (f), the particle undergoes more acceleration in the second island (Scenario 1). The particle then moves toward the right end of the third island without much acceleration during (g). After that, in (h) the particle gains more energy by scattering back into the outskirt of the third island, achieving further acceleration (Scenario 2).

\begin{figure}[ht!]
\centering
\includegraphics[width=\linewidth]{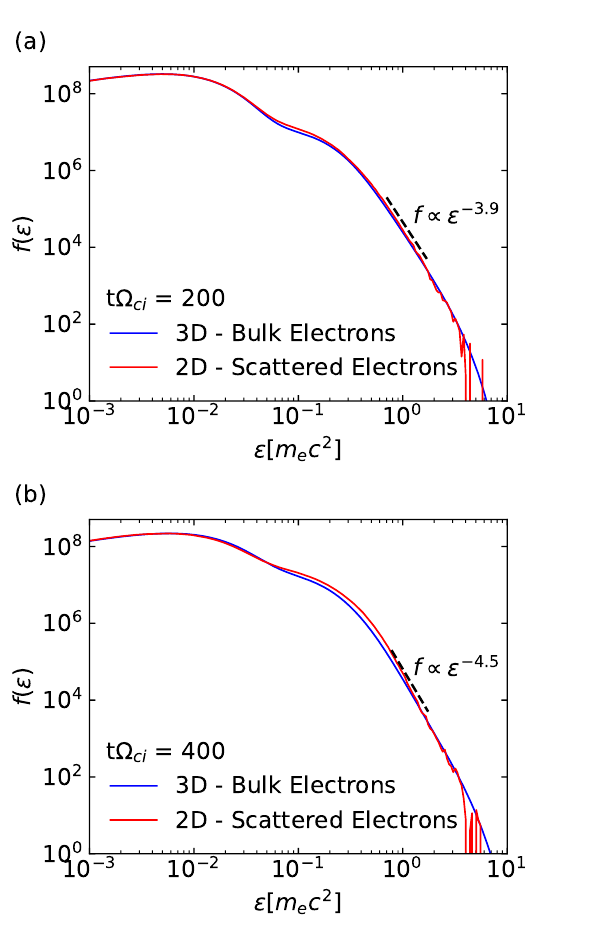}
\caption{Comparison between scattered 2D particle distribution with 3D results from \cite{Li2019} at $t\Omega_{ci} = 200$ and $400$, where the tracer species has an scattering frequency of $\nu = 0.1 \Omega_{ci}$ that has the closest match with the 3D spectrum. We normalize $f(\varepsilon)$ so that the electron numbers of both populations are the same. Results from the entire scan are shown in Figure \ref{fig:scattered_to_tracer_ratio}.
\label{fig:best_fit_e}}
\end{figure}

While pitch-angle scattering can enhance particle transport and acceleration in 2D simulations, it is unclear which scattering frequency is most likely to reproduce the 3D spectrum. To determine the optimal scattering frequency accurately, analysis of the particle transport coefficients (e.g., spatial, energy, and pitch-angle diffusion coefficients) in both the 2D and 3D simulations is necessary. However, such a study is beyond the scope of this paper since it requires a better understanding of the turbulence/fluctuation properties, an area still under active research~\citep{Daughton2011Role,Huang2016Turbulent,Kowal2017Statistics,Guo20213D}. Instead, here we attempt to find closest matches with 3D spectra by scanning the scattering frequency. The results show that the optimal $\nu\approx 0.08-0.2 \Omega_{ci}$ for our simulation parameters. Figure \ref{fig:best_fit_e} compares the scattered 2D electron spectrum with the 3D results at $t\Omega_{ci}=200$ and 400  when $\nu = 0.1 \Omega_{ci}$.
At both time frames, the spectral indices and maximum energies match well between the 2D and 3D simulations, where power-law indices are 3.9 and 4.5 in 3D, and 4.0 and 5.0 in 2D.

\begin{figure*}[ht!]
\centering
\includegraphics[width=\linewidth]{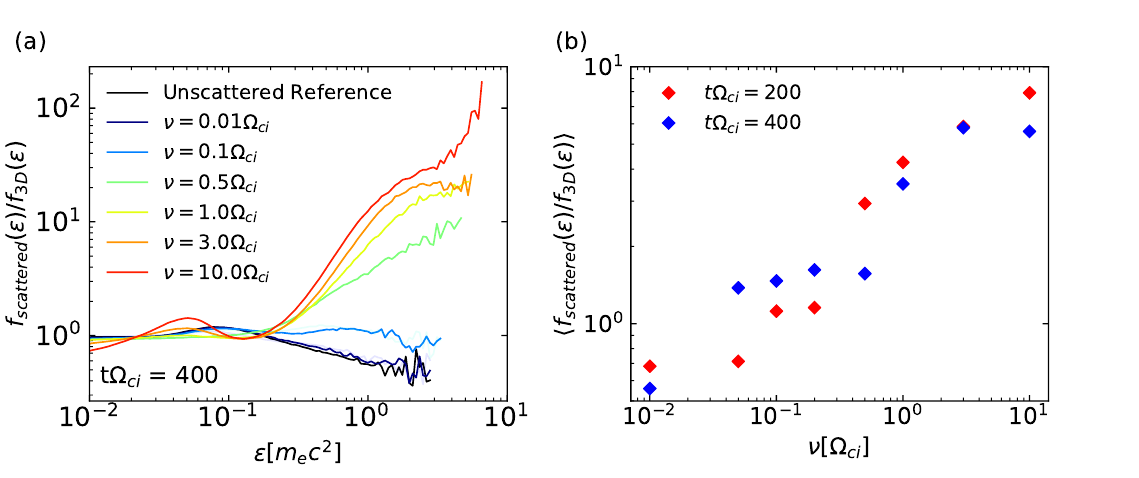}
\caption{(a) The spectrum ratios between the scattered particle species for different scattering frequencies, where the y-axis value is the 2D spectrum divided by the corresponding 3D spectrum. Lighter background lines are additional ratios with scattering frequencies between $\nu = 0-10 \Omega_{ci}$. Panel (b) includes all of these runs and displays the ratio of the integrated high-energy tail from $\varepsilon/m_ec^2 =$ 0.5 to 3.0 at $t\Omega_{ci}$ = 200, 400.
\label{fig:scattered_to_tracer_ratio}}
\end{figure*}

An alternative way to look at the spectral difference of 2D and 3D is to take their ratio as a function of energy, shown in Figure \ref{fig:scattered_to_tracer_ratio} (a). Here we plot an unscattered case as well as a scan over a wide range of scattering frequencies. At low frequencies, the particle spectrum of the scattered species remains relatively unchanged between the unscattered and scattered species. Only when the scattering grows to $\nu \gtrsim 0.05 \Omega_{ci}$ does the tail of the distribution in 2D begin to gain comparable energy to that of 3D. As the frequency increases further, however, we notice a deficit in lowest and mid-energy ranges and an overshoot in the tail compare with that of 3D, indicating too high a scattering frequency at this point. For further comparison, it is informative to examine panel (b), which shows the ratio of particle numbers in the energy range $0.5<\varepsilon/m_ec^2 <3.0$ in 2D and 3D simulations. This measure emphasizes the tail. It shows that the ratio increases approximately linearly from frequencies between $\nu \sim 0.05 - 1.0 \Omega_{ci}$ at $t\Omega_{ci}$ = 200, and becomes gradually saturated as $\nu$ further increases. As a retrospect, the dependence on the scattering frequency can be understood as the following. On the one hand, when the scattering frequency is too low, the particles do not gain significant additional Fermi acceleration since they are unable to diffuse significantly across field lines and leave their initial islands. When the scattering frequency is too high, the diffusion may take a significant fraction of particles away from the acceleration region, even into the upstream region. At $t\Omega_{ci}$ = 400, we see the same general trend at high and low scattering frequencies. Between $\nu \sim 0.05 - 0.5 \Omega_{ci}$, the dependence is flatted relative to $t\Omega_{ci}$ = 200. More analyses suggest that this is associated with variations in the amount of magnetic energy converted into the particles between simulations. For cases between $\nu = 0.05-0.2\Omega_{ci}$, around $30\%$ of the initial magnetic energy is converted into kinetic energy, while for $\nu = 0.5\Omega_{ci}$ only $\sim 23\%$ is converted by $t\Omega_{ci}$ = 400. This difference shows up as a depletion in the tail around $\varepsilon/m_ec^2 = 0.1$ to $1.0$ seen in panel (a). 

\begin{figure*}[ht!]
\centering
\includegraphics[width=\linewidth]{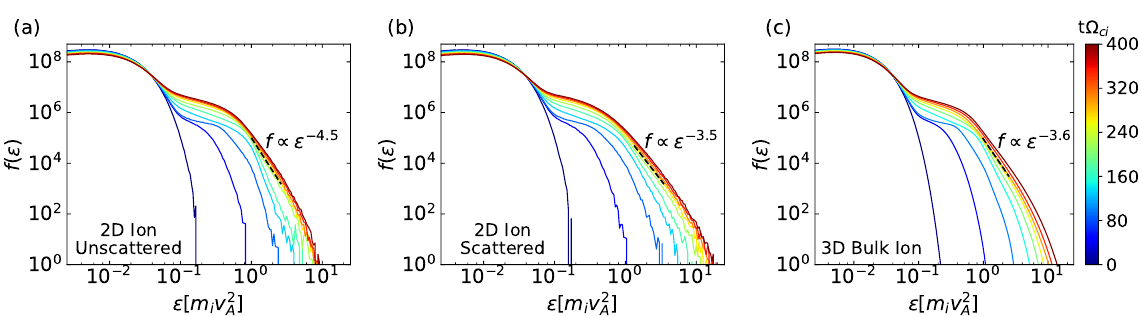}
\caption{Comparison of the ion spectrum between 2D and 3D at the same parameters as \citet{Zhang2021}. Since both the 2D and 3D ion spectrum begins to curve at the end of the simulations due to the periodic boundary conditions, slopes are measured at $t\Omega_{ci} = 240$ where the power-law is most distinguishable.
\label{fig:ion_spec}}
\end{figure*}

A final comparison is made with the 2D ion spectra to the 3D ion spectra from \citet{Zhang2021}, shown in Figure \ref{fig:ion_spec}. To make this comparison we double the domain size and number of grid points to match with the 3D simulation.  Using a scattering frequency for the scattered ion species of 0.04 $\Omega_{ci}$ we can closely compare the 2D scattered ion spectra to the unscattered and 3D bulk spectra, with a scattering frequency smaller than what is needed for electrons. This is likely due to the differences in gyroradii between ions and electrons. The spectral index was taken at the spectra corresponding to  $t\Omega_{ci} \sim 240$  before the 2D spectra began curving near the end of the simulation. While these 2D simulations appear to slightly overpopulate the very high energy particles at $\varepsilon > 5 m_i V_A^2$ (panel (b)), the agreement is fairly good.


\section{Conclusion} \label{sec:Conclusion}
Due to the artificial restriction of charged particles in 2D PIC simulations, the acceleration of particles is suppressed as they are buried inside inactive magnetic islands. This has led to discrepancies between past 2D and 3D reconnection simulations which we have highlighted in their energy spectra.
Here we have introduced ad hoc pitch-angle scattering in 2D simulations to break this limitation and diffuse particles from their initial field lines. Our surveys over different scattering frequencies demonstrates its impact on scattered electron and ion distributions. We have demonstrated that this can bring 2D results close to 3D results where the particle transport is facilitated by turbulent field lines \citep{Dahlin2017,Li2019,Zhang2021}. More specifically, we have shown that our simple pitch-angle scattering model can reproduce the several characteristic features missed in earlier 2D simulations, allowing the particles to leave magnetic islands, hardens the spectrum, and increasing the cutoff energy. The more efficient acceleration is accomplished by breaking the particles' tie to the field lines and allowing the scattered particles to access two scenarios. First is that some particles transport to other islands. The second way is that some particles can achieve higher energy if they are scattered back to the outskirt of islands (the exhaust regions) and experiences more acceleration.

Ad hoc pitch-angle scattering opens the possibility of mimicking the 3D effects (efficient transport, hardening of the spectrum, and increasing the high energy cutoff) without the expense of running full 3D simulations. Our model may be improved by further refinements on the scattering model which would result in higher fidelity 2D simulations. 
Additional future work in refining the 3D particle transport in the reconnection layer is essential. More accurate functions for the scattering frequency could be found from 3D simulations, and used to build a more accurate transport model. While here we envision transport due to field line chaos and pitch-angle scattering in magnetic turbulence, theory that includes the diffusion of particles due to magnetic islands \cite{Zank2014,leRoux2015,le2018self} may provide additional basis, where the analytical diffusion coefficients have been derived. These could provide a more generally applicable model that encapsulates the essential physics, and help achieve a more complete picture for particle acceleration in magnetic reconnection. 

\begin{acknowledgments}
\begin{nolinenumbers}
We greatly appreciate the discussion with Dr. Qile Zhang, and the 3D data he provided for the comparison with our 2D results. 
This material is based upon work supported by the U.S. Department of Energy, Office of Science, Office of Advanced Scientific Computing Research, Department of Energy Computational Science Graduate Fellowship under Award Number DE-SC0021110. This research was in part supported by the Los Alamos National Laboratory (LANL) through its Center for Space and Earth Science
(CSES) and LDRD program. CSES is funded by LANL’s Laboratory Directed Research and Development (LDRD) program under project number 20210528CR. F.G. also acknowledge support in part from NASA Grant 80HQTR20T0073, 80HQTR21T0087 and 80HQTR20T0040, 80HQTR21T0104 and DOE grant DE-SC0020219.  X.L. acknowledges the support from NASA through Grant 80NSSC21K1313, National Science Foundation Grant No. AST-2107745, and Los Alamos National Laboratory through subcontract No. 622828. This research used resources of the National Energy Research Scientific Computing Center (NERSC), a U.S. Department of Energy Office of Science User Facility located at Lawrence Berkeley National Laboratory, operated under Contract No. DE-AC02-05CH11231. Additional simulations were performed at the Texas Advanced
Computing Center (TACC) at The University
of Texas at Austin and LANL Institutional Computing Resource.  
\end{nolinenumbers}
\end{acknowledgments}

\bibliography{paper_bib}{}
\bibliographystyle{aasjournal}

\end{document}